\documentclass[journal]{IEEEtran}
\usepackage{amsmath,amsfonts}
\usepackage{algorithmic}
\usepackage{algorithm}
\usepackage{array}
\usepackage[caption=false,font=normalsize,labelfont=sf,textfont=sf]{subfig}
\usepackage{textcomp}
\usepackage{stfloats}
\usepackage{url}
\usepackage{verbatim}
\usepackage{graphicx}
\usepackage{cite}
\usepackage{hyperref}
\usepackage{multirow}
\usepackage{multicol}
\usepackage{makecell}
\usepackage{diagbox}
\usepackage{balance}
\usepackage{color,soul}
\usepackage{bbding}
\usepackage{ctable} 
\usepackage{tabu}
\usepackage{amssymb}
\usepackage{pifont}
\usepackage{flushend} 

\DeclareMathOperator*{\argmax}{arg\,max}

\begin{document}

\title{Towards Decoupling Frontend Enhancement and Backend Recognition in Monaural Robust ASR}

\author{Yufeng Yang,~\IEEEmembership{Student Member,~IEEE,} Ashutosh Pandey,~\IEEEmembership{Member,~IEEE}, \\and DeLiang Wang,~\IEEEmembership{Fellow,~IEEE}
\thanks{This research was supported by an NIH grant (R01DC012048), the Ohio Supercomputer Center, the Pittsburgh Supercomputer Center (NSF ACI-1928147), and the NCSA Delta System (NSF OAC-2005572).}
\thanks{Yufeng Yang and Ashutosh Pandey are with the Department of Computer Science and Engineering, The Ohio State University, Columbus, OH, 43210 USA (email:\href{mailto:yang.5662@osu.edu}{yang.5662@osu.edu}; \href{mailto:pandey.99@osu.edu}{pandey.99@osu.edu})}

\thanks{DeLiang Wang is with the Department of Computer Science and Engineering, and the Center for Cognitive and Brain Sciences, The Ohio State University, Columbus, OH, 43210 USA (email: \href{mailto:dwang@cse.ohio-state.edu}{dwang@cse.ohio-state.edu})}
}




\maketitle

\begin{abstract}
It has been shown that the intelligibility of noisy speech can be improved by speech enhancement (SE) algorithms. However, monaural SE has not been established as an effective frontend for automatic speech recognition (ASR) in noisy conditions compared to an ASR model trained on noisy speech directly. The divide between SE and ASR impedes the progress of robust ASR systems, especially as SE has made major advances in recent years. This paper focuses on eliminating this divide with an ARN (attentive recurrent network) time-domain and a CrossNet time-frequency domain enhancement models. The proposed systems fully decouple frontend enhancement and backend ASR trained only on clean speech. Results on the WSJ, CHiME-2, LibriSpeech, and CHiME-4 corpora demonstrate that ARN and CrossNet enhanced speech both translate to improved ASR results in noisy and reverberant environments, and generalize well to real acoustic scenarios. The proposed system outperforms the baselines trained on corrupted speech directly. Furthermore, it cuts the previous best word error rate (WER) on CHiME-2 by $28.4\%$ relatively with a $5.57\%$ WER, and achieves $3.32/4.44\%$ WER on single-channel CHiME-4 simulated/real test data without training on CHiME-4.

\end{abstract}


\section{Introduction}
\IEEEPARstart{I}{n} real environments, acoustic interference is ubiquitous in speech communication, and negatively impacts the performance of speech-based applications such as smart home devices \cite{heymann2018performance} and conference transcription systems \cite{fu2021aishell}. To attenuate acoustic interference, speech enhancement (SE) algorithms estimate clean speech from noisy or reverberant speech. These algorithms have achieved remarkable success in improving the quality and intelligibility of speech with background interference, particularly since the introduction of deep learning to the field \cite{wang2018supervised}. However, a major disappointment is that monaurally enhanced speech does not translate to improved automatic speech recognition (ASR), and this has been attributed to the distortion introduced by monaural SE \cite{wang_bridging_2019}. The divide between SE and ASR has persisted despite considerable research over decades to bridge enhancement and ASR \cite{cooke2001robust, raj2004reconstruction, narayanan2014investigation, wang_bridging_2019, zhang2021closing}. This study represents a new effort to bridge the fields of SE and ASR.

Traditional SE approaches are based on spectral subtraction, Wiener filtering, and statistical model based approaches \cite{loizou2007speech}. SE was first formulated as a deep learning problem in \cite{wang2013towards}, and this formulation has led to remarkable advances in recent years \cite{wang2018supervised}. In this formulation, SE is mainly conducted in the time-frequency (T-F) domain, such as short-time Fourier transform (STFT). Common practice is to enhance spectral magnitudes and reconstruct the enhanced waveform using the enhanced magnitude and noisy phase. However, this approach ignores phase information, which is important for enhancement performance \cite{paliwal2011importance}. A recent trend is to enhance both spectral magnitude and phase in approaches such as complex ratio masking \cite{williamson2015complex, hu2020dccrn}, which estimates the complex ideal ratio mask by predicting the real and imaginary spectrograms. Alternatively, complex spectral mapping predicts directly the real and imaginary spectrograms of clean speech from the complex STFT of noisy speech \cite{fu2017complex, tan2019learning, wang2020complex}. Different from spectral methods, time-domain enhancement predicts clean speech samples from noisy speech samples, hence enhancing speech magnitude and phase simultaneously \cite{fu2017raw, pascual2017segan, luo2019conv}. In this study, we select a time domain and a T-F domain SE model as frontends. For time-domain SE, we employ the attentive recurrent network (ARN), which incorporates an RNN (recurrent neural network) with a self-attention mechanism \cite{pandey2022self}. For T-F domain SE, a recently proposed CrossNet is employed, and it includes a cross-temporal module in addition to narrow-band and cross-band modules \cite{vahid2024crossnet}{. Both ARN and CrossNet achieve excellent enhancement performance.}



In terms of robust ASR, prevailing approaches perform acoustic modeling directly on noisy speech for noise-dependent or noise-independent models, which is proven to be effective on series of CHiME challenges \cite{barker2013pascal, vincent2013second, barker2015third, vincent2017analysis}. The drawback of such approaches is that noise-dependent models do not generalize well to untrained noises, and noise-independent models need an enormous amount of transcribed noisy speech for training, which is not only costly but also infeasible in many real-world applications. When tested on clean speech, an ASR model trained on noisy speech also results in an unavoidable performance gap compared with a corresponding model trained on clean speech, due to the mismatch between training and test conditions. 

To bridge the divide between SE and ASR, attempts have been made to perform acoustic modeling on enhanced speech \cite{seltzer2013investigation}, \cite{meng2018adversarial}, \cite{ wang_bridging_2019} or enhanced features \cite{wang_enhanced_2019} in a distortion-independent way. However, the backends of these systems are dependent on the frontends. When a frontend is replaced, say by a better enhancement model, the backend would need to be retrained or ASR performance degrades. Conversely, SE can be designed to serve ASR. In \cite{plantinga2021perceptual}, a perceptual loss based model was proposed to guide the training of a SE frontend using senone labels from an acoustic model (AM). In \cite{ho2023naaloss}, a noise- and artifact-aware loss function is designed to train the frontend. The study in \cite{narayanan2023learning} focuses on improving T-F mask quality. In \cite{iwamoto2022bad, zorilua2022speaker}, mixing enhanced speech with noisy speech mitigates processing artifacts and improves ASR, but the mixed speech is still noisy. In \cite{kinoshita2020improving}, SE is shown as an effective frontend for ASR, but the backend is trained on multi-conditioned speech, mismatched with the enhanced speech which is aimed to be clean. To combine SE and ASR, a SE model and an ASR model can be jointly trained \cite{xu2019joint, menne2019investigation, zhu2022joint}. Similar ideas are also utilized in end-to-end (E2E) systems \cite{shi2022train, chang2022end}. Such systems often have a huge model size and are difficult to train. Furthermore, the frontend and backend inside a joint or E2E system are dependent on each other, making it problematic to benefit from an improved frontend or backend individually, hence limiting the flexibility in real applications. Results in \cite{masuyama2023exploring} demonstrate that, although fine-tuning an E2E system lowers word error rate (WER), the individual performance of the frontend degrades significantly when it is decoupled after joint fine-tuning.

In this paper, we investigate the most straightforward approach to robust ASR where an ASR model is trained on clean speech only and its input is frontend enhanced speech directly. Thus the proposed robust ASR system completely decouples the frontend and backend. In other words, the frontend performs SE without ASR considerations, and the backend ASR model is designed to recognize clean speech without considering the potential distortion of an enhancement frontend. By combining two modules directly to recognize noisy speech, we demonstrate that the proposed system outperforms alternative robust ASR systems, including noise-independent, reverberation-independent, and interference-independent models, on noisy, reverberant, and reverberant-noisy speech. When tested on the medium vocabulary track (track 2) of the CHiME-2 corpus, our system achieves a $5.57\%$ average WER. To our knowledge, this result represents the best on this dataset to date and outperforms the previous best by $28.4\%$. Results on the Wall Street Journal (WSJ) and LibriSpeech corpora show that the proposed system outperforms all baselines in noisy and reverberant conditions. Our system also generalizes well to CHiME-4 using an off-the-shelf frontend trained on LibriSpeech and a backend trained on WSJ clean speech, and achieves $3.32/4.44\%$ WER on single-channel CHiME-4 simulated/real test data without training on CHiME-4. Our investigation reveals that using short-time objective intelligibility (STOI) \cite{taal2011algorithm} as the model selection criterion is superior for SE models in terms of ASR. Our study makes three main contributions to monaural robust ASR. First, we advocate the approach of decoupling a frontend devoted to enhancing noisy speech and a backend devoted to recognizing clean speech, which contrasts with the prevailing approach that trains ASR on noisy speech. Second, we demonstrate that the decoupled approach outperforms the prevailing approach on multiple datasets. Third, we advance the state-of-the-art ASR results on CHiME-2 and CHiME-4 datasets. Unlike the previous best systems on single-channel CHiME-4, we obtain the top performance with a pretrained speech enhancement frontend.

This paper expands a preliminary study \cite{yang2022time} in several major ways. First, on the task of recognizing noisy speech, we investigate the proposed system on the WSJ, LibriSpeech, and CHiME-4 corpora in addition to CHiME-2, introducing new test conditions and speech materials. Second, we evaluate the proposed system in reverberant and reverberant-noisy conditions. Third, we add another noise-independent baseline trained on more data than the noise-independent model in \cite{yang2022time}. Fourth, we extend the AM from a hybrid system of deep neural network and hidden Markov model (DNN-HMM) in \cite{yang2022time} to an E2E architecture. Fifth, we compare the decoupled ASR system with two more baseline ASR systems on the LibriSpeech corpus trained on dynamically generated noisy speech and multi-conditioned speech, matching the data seen by the SE frontend during training, further demonstrating the effectiveness of the proposed decoupled system. Sixth, we add experiments on CHiME-2 with a T-F domain CrossNet frontend, and demonstrate that decoupling the frontend and backend is not limited to time-domain SE. Last, the decoupled system achieves a $3.32/4.44\%$ WER on the single-channel CHiME-4 simulated/real test sets, demonstrating strong generalization ability to cross-corpus real acoustic scenarios.

The remainder of the paper is organized as follows. Section~\ref{sec:system} describes the frontend and backend models for SE and ASR. Section~\ref{sec:exp} describes the experimental setup and implementation details. Evaluation results and comparisons are presented in Section~\ref{sec:result}. Section~\ref{sec:conclusion} concludes the paper.

\section{System Description}\label{sec:system}

\subsection{Problem Formulation}

\subsubsection{Monaural Speech Enhancement}
The monaural SE problem can be formulated as follows:

\begin{equation}\label{eq:se}
\begin{aligned}
\mathbf{y} = \mathbf{h}*\mathbf{s} + \mathbf{n},
\end{aligned}
\end{equation}
where $\mathbf{y}$, $\mathbf{h}$, $\mathbf{s}$, and $\mathbf{n}$ denote noisy speech, room impulse response (RIR), clean speech, and additive noise, respectively; $*$ denotes convolution. SE computes an estimate of $\mathbf{s}$, $\hat{\mathbf{s}}$, from $\mathbf{y}$. When the speech signal is anechoic, $\mathbf{h}$ can be omitted. When the signal is reverberant, Eq.~\ref{eq:se} can be expressed in another way as

\begin{equation}
\begin{aligned}
\mathbf{y}&= (\mathbf{h}_{d} + \mathbf{h}_{r}) * \mathbf{s} + \mathbf{n} \\
&=\mathbf{h}_d * \mathbf{s} +  \mathbf{h}_{r} * \mathbf{s} +\mathbf{n} \\
&= \mathbf{s}_d + \mathbf{s}_{r} + \mathbf{n},
\end{aligned}
\end{equation}
where $\mathbf{s}_d$ and $\mathbf{s}_{r}$ denote the direct-path and reverberated speech, respectively. The RIRs of direct-path speech and reverberated speech are denoted by $\mathbf{h}_d$ and $\mathbf{h}_{r}$, respectively. SE aims to estimate $\mathbf{s}_d$ in this study.

\subsubsection{Automatic Speech Recognition}
An ASR system computes the optimal word sequence $\mathbf{W^{*}}$ given a sequence of acoustic features $\mathbf{X}$ of speech signal $\mathbf{x}$, which is formulated as a maximum \emph{a posteriori} probability problem

\begin{equation}\label{eq:map}
    \mathbf{W^{*}} = \argmax_{\mathbf{W}} P_{\mathcal{AM}, \mathcal{LM}}(\mathbf{W} | \mathbf{X}),
\end{equation}
where $\mathcal{AM}$ and $\mathcal{LM}$ denote an AM and language model (LM), respectively. Using Bayes' theorem, Eq.~\ref{eq:map} can be written as

\begin{equation}
    \mathbf{W^{*}} = \argmax_{\mathbf{W}} p_{\mathcal{AM}}(\mathbf{X} | \mathbf{W})P_{\mathcal{LM}}(\mathbf{W}),
\end{equation}
where $p_{\mathcal{AM}}$ and $P_{\mathcal{LM}}$ are AM likelihood and LM prior probability, respectively. An AM predicts the likelihood of acoustic features of a phoneme or another linguistic unit, and an LM provides a probability distribution over words or sequences of words in a speech corpus. In an E2E ASR system, a word sequence is predicted directly given $\mathbf{X}$.

\subsection{Frontend Networks}
\subsubsection{Attentive Recurrent Network}
We employ ARN as the time-domain frontend of ASR, which comprises RNN, self-attention, feedforward network, and layer normalization modules. Details of ARN building blocks can be found in \cite{pandey2022self}. In this work, the non-causal version of ARN is used, namely the RNN in ARN is bi-directional long short-term memory (BLSTM) and self-attention is unmasked.



\begin{figure}[htbp!]
    \centering
    \includegraphics[width=\linewidth]{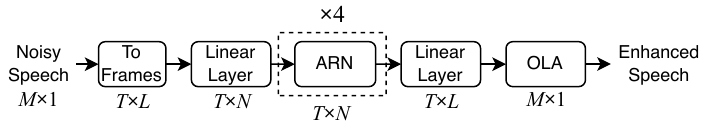}
    \caption{Diagram of ARN for speech enhancement. $T$ is the total number of frames and $L$ is the frame length.}
    \label{fig:arn}
\end{figure}

Fig.~\ref{fig:arn} shows the diagram of ARN. After an input signal with $M$ samples is chunked into overlapping frames, all frames are projected into a latent representation of size $N$ by a linear layer. Then the latent representation is processed by four consecutive ARN blocks, and another linear layer projects the output of the last ARN block back to size $L$. The enhanced speech is finally computed using the overlap-and-add (OLA) method.

\subsubsection{CrossNet}
We use CrossNet as the T-F domain frontend for monaural SE \cite{vahid2024crossnet}. CrossNet is motivated by SpatialNet \cite{quan2023spatialnet}, but introduces a cross-temporal module after the cross-band module and positional encoding. This modification enhances temporal processing for monaural separation. CrossNet performs complex spectral mapping by predicting the real and imaginary (RI) parts of the STFT of clean speech from the stacked RI parts of the STFT of noisy speech \cite{williamson2015complex, tan2019learning}. The enhanced waveform is generated by performing an inverse STFT on the enhanced RI parts.


\subsection{Backend Networks}
\subsubsection{Wide Residual Conformer Acoustic Model}
\begin{figure}[htbp!]
    \centering
    \includegraphics[width=\linewidth]{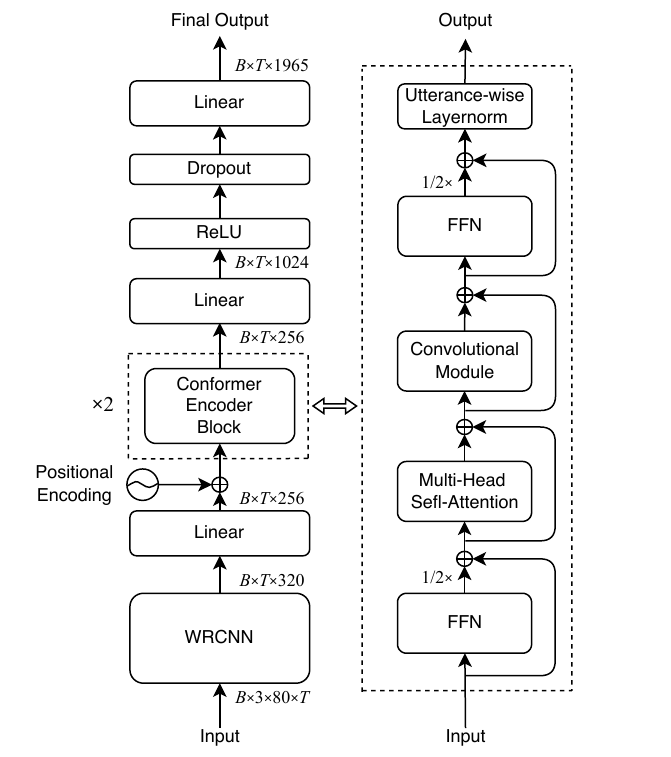}
    \caption{System architecture of a WRConformer AM. $B$ denotes the batch size, and $T$ denotes the number of time frames of the longest utterance in a batch.}
    \label{fig:cam}
\end{figure}

We utilize a Conformer-based AM \cite{yang2022conformer} as the backend of the proposed system, denoted as WRConformer AM. It is built upon a wide residual BLSTM network (WRBN) (see \cite{wang_bridging_2019}) and shown to outperform it on the CHiME-4 single-channel track \cite{yang2022conformer}. The system architecture of WRConformer AM is shown in Fig.~\ref{fig:cam}, where FFN denotes a feedforward network. WRConformer AM takes as input 80-dimensional mean-normalized log-Mel filterbank features extracted from the frontend output, coupled with its delta and delta-delta features. First, the input is processed by a wide residual convolutional layer denoted as WRCNN, which passes the input signal through a convolution layer and uses three residual blocks to extract representations at different frequency resolutions \cite{zagoruyko_wide_2016}. Afterwards, an utterance-wise batch normalization and a linear layer with ELU (exponential linear unit) non-linearity are utilized to project the signal onto 320 dimensions. Then a linear layer projects the signal onto the dimension of multi-head self-attention. After processing by two blocks of the Conformer encoder with absolute positional encoding, the signal is projected onto 1024 dimensions, followed by ReLU (rectified linear unit) activation and dropout. Next, a linear layer projects the signal to the final output of each frame as the posterior probability of 1965 context-dependent senone states. Then the output is sent to a decoder for text transcripts.

\subsubsection{End-to-end Wide Residual Conformer Network}\label{sec:e2e_conformer}
We extend WRConformer AM into a connectionist temporal classification (CTC) and attention Conformer-encoder Transformer-decoder E2E ASR model. Leveraging the ASR recipe implemented in ESPnet \cite{watanabe2018espnet}, we adapt the standard CTC/attention Conformer-encoder Transformer-decoder ASR recipe to E2E WRConformer. In this adaptation, we replace the 2-D convolution in the subsampling module with modified WRCNN, which comprises two ResBlocks (see \cite{jahn2016wide}). The first ResBlock projects an input log-Mel feature to 512 dimensions, while the second Resblock maintains the same input and output dimensions. Each ResBlock subsamples time frames by a factor of 2, resulting in the total number of frames reduced by a factor of 4 after the processing of the subsampling module, matching the default ESPnet subsampling module. In E2E WRConformer, the number of Conformer encoders is set to 10 and the other configurations are the same as the default setting. The implementation is available online \footnote{\href{https://github.com/yfyangseu/espnet}{https://github.com/yfyangseu/espnet}}.

\section{Experimental Setup}\label{sec:exp}

\subsection{Frontend Enhancement}
\subsubsection{Reverberation Generation}
RIRs are generated in the same way as \cite{pandey2022tparn}. Room length and width range in [5, 10] m, and height in [3, 4] m. A sound source and a microphone are sampled at least 0.5 m away from the walls. The distance between the two ranges from 0.75 m to 2 m. We use Pyroomacoustics \cite{scheibler2018pyroomacoustics} for RIR generation, which is a hybrid method where the image method with an order of 6 is used to model early reflections and late reverberation is modeled by ray-tracing. RIRs for training, validation, and test data are randomly sampled from 319666, 40091, and 40243 generated RIRs, respectively.

\subsubsection{WSJ}\label{sec:data_wsj}
We utilize the WSJ0 SI-84 corpus \cite{paul1992wsj} to train ARN for SE and WRConformer AM for ASR. This dataset comprises English utterances from 42 male and 41 female speakers, providing 7138, 1206, and 330 utterances for training, validation, and test, respectively. For ARN training on WSJ, we design three ARN systems aiming at denoising, dereverberation, or both, denoted as DN-ARN, DR-ARN, and NR-ARN, respectively. 

Training data for DN-ARN is generated by randomly mixing 7138 anechoic-clean WSJ training utterances with noise segments randomly picked from the 10k non-speech sounds of a sound effect library (\href{http://www.sound-ideas.com}{http://www.sound-ideas.com}) with signal-to-noise ratio (SNR) uniformly selected in [$-7$, $0$] dB and [$0$, $10$] dB ranges, with $50\%$ probability for each range. Validation data is generated by mixing 1206 anechoic-clean WSJ validation utterances with the factory noise from the NOISEX-92 dataset \cite{varga1993assessment} at $-6$ dB SNR. Once trained, DN-ARN is tested on noisy speech generated by mixing 330 anechoic-clean WSJ test utterances with ADTBabble and ADTCafeteria noises (\href{http://www.auditec.com/}{http://www.auditec.com/}) at \{-6, -3, 0, 3, 6, 9\} dB SNR levels.

Training data for DR-ARN is generated by randomly convolving 7138 anechoic-clean WSJ training utterances with training RIRs in the T60 (room reverberation time) range of [0.2, 1.0] s. DR-ARN is validated on reverberant-clean utterances generated by convolving 1206 anechoic-clean WSJ validation utterances with validation RIRs in the T60 range of [0.8, 1.0] s. Once trained, DR-ARN is evaluated on reverberant speech generated by convolving 330 anechoic-clean WSJ test utterances with test RIRs in the T60 range of [0.2, 0.4] s, [0.4, 0.6] s, [0.6, 0.8] s, and [0.8, 1.0] s separately. The training target of DR-ARN is direct-path speech, which is generated by convolving the anechoic-clean speech with the direct-path RIR. When generating a test utterance, an anechoic-clean speech signal with $L_{s}$ samples is convolved with a direct-path RIR whose peak value is at the $p$-th sample to generate reverberant-clean speech. Then the first $p - 1$ samples of the reverberant-clean speech signal are removed and the next $L_{s}$ samples are retained for testing. This technique addresses the alignment issue for ASR decoding, such that the direct-path speech signals of different RIRs have the same WER when tested on the backend trained on anechoic-clean speech.

Training data for NR-ARN is generated by randomly convolving 7138 anechoic-clean WSJ training utterances with training RIRs in the same way as for DR-ARN and then mixing with 10k noises in the same way as for DN-ARN. Validation data is generated in the same way as that for DR-ARN but mixed with factory noise at $-6$ dB SNR. Test data for NR-ARN is generated in the same way as for DR-ARN but mixed with noise in the same way as for DN-ARN such that test speech is both reverberant and noisy. For better comparison, noise segments added to the reverberant-clean speech for each level of T60 are the same.

DN-ARN, DR-ARN, and NR-ARN share the same model architecture. The sampling rate for all utterances is 16 kHz. All training samples are generated randomly and dynamically for all ARNs. We apply root mean square normalization to noisy mixtures, and clean speech is scaled to produce a specified SNR level. During training, the number of samples for each utterance is set to 64000. Input and output frame size is set to 16 ms with a 2 ms frame shift. Dimension $N$ for BLSTM is set to 1024 such that forward and backward hidden states are sized at 512. The dropout rate is set to $0.05$ for feedforward blocks. The ARNs are trained using the PCM (phase-constrained magnitude) loss \cite{pandey2021dense}.




One epoch of training consists of 157036 utterances, and the ARNs are trained for 100 epochs with batch size 16. The Adam optimizer \cite{kingma2015adam} is utilized. The learning rate of the first 33 epochs is fixed to $2e^{-4}$ and then exponentially decays every epoch till the final learning rate of $2e^{-5}$. All ARNs are trained on two NVIDIA V100 GPUs. Because STOI is shown to relate to WER \cite{moore2017speech}, we also use validation STOI as a model selection criterion in addition to validation PCM loss. Validation STOI is computed by averaging the STOI scores on all utterances in the validation set for each epoch. Once the training is done, the checkpoint corresponding to the best validation score is selected.

\subsubsection{CHiME-2}\label{sec:data_chime2}
Our experiments are conducted on the medium vocabulary track (track 2) of the CHiME-2 corpus \cite{vincent2013second}. It is a commonly used dataset for robust ASR evaluation and is generated by convolving WSJ clean speech with binaural RIRs (BRIRs) and mixing with non-stationary family home noise \cite{barker2013pascal}. Although the speech materials are from the WSJ corpus, due to the BRIRs used in the data generation process \cite{christensen2010chime}, speech signal alignments are altered, which makes it infeasible to use WSJ anechoic-clean speech as the training target for SE. Therefore, we treat reverberant-clean speech as the target signal for training DN-ARN. Two channels are averaged to produce single-channel speech. We employ ARN for time-domain and CrossNet for T-F domain SE on denoising.

Training data for DN-ARN and CrossNet is generated by randomly mixing 7138 reverberant-clean training utterances from CHiME-2 with 10k noises in the same way as for DN-ARN for WSJ. Validation data is generated by mixing 409 reverberant-clean validation utterances from CHiME-2 with the factory noise at $-6$ dB SNR. Once trained, the frontends are tested on reverberant-noisy test data, which has six SNR levels with each containing $330$ utterances. The training setup of DN-ARN on CHiME-2 is the same as DN-ARN on WSJ.

For CrossNet, the network configurations are kept the same as in \cite{vahid2024crossnet}. The learning rate schedule and batch sizes are the same as those of DN-ARN trained on WSJ. One CrossNet is trained with the PCM loss, and the other is trained with scale-invariant signal-to-distortion ratio (SI-SDR) \cite{leroux2019sdr} training and validation loss.

\subsubsection{LibriSpeech}\label{sec:exp_se_libri}

We evaluate the denoising performance of ARN on the LibriSpeech corpus \cite{panayotov2015librispeech}, which contains around 1000 hours of read English speech, sourced from audiobooks available in the public domain, through the LibriVox project \footnote{\href{https://librivox.org}{https://librivox.org}}. LibriSpeech has different speech materials from WSJ or CHiME-2. The training data is generated by mixing all 960 hr training utterances from train-clean-100, train-clean-360, and train-other-500, with 10k noises at SNR uniformly selected within the [-5, 0] dB and [0, 10] dB ranges, with $50\%$ probability for each range. Validation data is generated by mixing factory noise with all dev-clean and dev-other utterances at $-5$ dB SNR. The test-clean and test-other utterances are mixed with ADTBabble and ADTCafeteria noise at \{-5, -2, 0, 2, 5, 10\} dB SNR levels for evaluation.

DN-ARN on LibriSpeech is trained on two NVIDIA A100 GPUs, with a batch size of 32. The learning rate schedule, optimizer, and loss function are the same as those of DN-ARN on WSJ.

The SE performance is evaluated using standard STOI and perceptual evaluation of speech quality (PESQ) \cite{rix2001pesq} metrics. STOI ranges typically between [$0$, $1$] and indicates speech intelligibility, usually in percentage. PESQ ranges between [$-0.5$, $4.5$] and a higher score denotes higher speech quality.

\subsection{Backend Recognition}

\subsubsection{WSJ}
The backend of the proposed DN-ARN, DR-ARN, and NR-ARN systems is the WRConformer AM trained on anechoic-clean WSJ speech, with 7138, 1206, and 330 utterances for training, validation, and test, respectively.

All WRConformer AMs share the same model architecture. The configuration of WRCNN is kept the same as in \cite{yang2022conformer}. The attention dimension is set to $256$. The kernel size of the $1$-D depthwise convolution is set to $16$. We use the learning rate schedule from \cite{vaswani2017attention}, with $2$k warm-up steps and a learning rate factor $k$ of $100$. The Adam optimizer with $\beta_{1}=0.9$, $\beta_{2}=0.98$, and $\epsilon=1e^{-9}$ is utilized for model training. The batch size is set to $3$ and short utterances are padded with zeros to match the length of the longest utterance in each batch. The dropout rate is set to $0.15$ for network and attention weights. All WRConformer AMs are trained for 25 epochs and model selection is based on the cross-entropy loss on the validation set. Training labels are 1965 senones in our experiments, and they are generated as in \cite{wang2016joint}. For log-Mel feature extraction, we follow the approach in \cite{wang_bridging_2019}, but skip pre-emphasizing, dithering, and direct-current offset removal steps. A Hamming window is applied to the input waveform for STFT. Then a small value of $e^{-40}$ is added to prevent the underflow of logarithmic operation.

The decoder is the same as in \cite{wang2016joint}. AM outputs are first subtracted by the log priors and then fed to the decoder, which is based on the CMU pronunciation dictionary and the official 5k close-vocabulary tri-gram language model. The decoding beamwidth is set to 13, and the lattice beamwidth is 8. The number of active tokens ranges from 200 to 700. Language model weights ranging from 4 to 25 are utilized.

We train two baseline WRConformer AMs for each system. One is trained on 157036 utterances (denoted as 160k) following the approach in \cite{wang_bridging_2019}, and another AM is trained on 320k utterances (denoted as 320k) in the same way for a potentially better baseline. Each baseline is trained using the same inputs as those to the corresponding frontend. That is, the baselines for DN-ARN, DR-ARN, and NR-ARN are trained on anechoic-noisy, reverberant-clean, and reverberant-noisy speech, respectively. The value of the learning rate factor $k$ for this model training is divided by $10$ after the $6$-th epoch and by another 10 after the $12$-th epoch. Decreasing the value of $k$ plays the role of fine-tuning and yields better training. Model trained on 320k reverberant-noisy speech for the DR-ARN system overfits during training, so $k$ is initially set to $50$ and divided by $10$ after the $12$-th epoch. We believe that the smaller variation of reverberation than additive noise accounts for the overfitting in the DR-ARN system.

To match the learning rate with the backend trained on 160k utterances, we modify the learning rate for the backend trained on 320k utterances. One learning rate value is used twice before changing to the next value in the learning rate schedule of the backend trained on 160k utterances. For example, if [$lr1$, $lr2$, $lr3$] are three consecutive learning rates for the backend trained on 160k utterances, the backend trained on 320k utterances will use the learning rates of [$lr1$, $lr1$, $lr2$, $lr2$, $lr3$, $lr3$] for six consecutive training steps. This modification makes the learning rates of the two backends epoch-wise consistent.

\subsubsection{CHiME-2}
The backend of the proposed DN-ARN system is the WRConformer AM trained and validated on CHiME-2 7138 and 409 reverberant-clean utterances, respectively. It has the same training setup with the backend on WSJ. We train two baseline WRConformer AMs on 160k and 320k utterances with the same settings as WSJ on reverberant-noisy speech. Following the official CHiME-2 recipe, We also train and validate a noise-dependent WRConformer AM (denoted as noise-dependent backend) on 7138 and 409 CHiME-2 reverberant-noisy utterances, respectively. It is trained with the same settings as other backends except for a learning rate factor $k$ of $1e^4$ and $5$k warm-up steps.

\subsubsection{LibriSpeech}
The backend utilized for LibriSpeech is the E2E WRConformer that has 10 Conformer encoders, 6 Transformer decoders, and an attention dimension of 512 with 8 attention heads. The feedforward layer operates with a dimension of 2048. A dropout rate of 0.1 is applied. STFT frame and shift sizes are 512 and 160, respectively. The CTC weight is set to 0.3, and the label smoothing weight is 0.1. The E2E WRConformer is trained for 50 epochs on the LibriSpeech 960 hr dataset on four NVIDIA A100 GPUs.

We create a training set by mixing the LibriSpeech 960 hr training set and 10k noises in the same way as for DN-ARN on LibriSpeech. To ensure robust learning, each utterance in the LibriSpeech training set is mixed with 4 randomly selected noise segments, resulting in a total of 3840 hr noisy speech (denoted as 4k hr). The validation set is the same as for DN-ARN on LibriSpeech.

\begin{table*}[!htbp]
    \centering
    \caption{ASR ($\%$WER) Results of the Proposed DN-ARN System and Comparison Systems on WSJ. `Enh.' Denotes `Enhanced'. All ASR Models Are Trained on Anechoic Speech.}
    \label{tab:asr_wsj_dnarn}
    \centering
    \scalebox{1}{
    \begin{tabular}[width=\linewidth]{ c l l   l l  c  c c c c c c }
        \specialrule{1pt}{0pt}{4pt}
         \multirow{2}{*}{\makecell[c]{Test\\ Noise}} & \multirow{2}{*}{Row} &\multirow{2}{*}{\makecell[c]{SE\\Network}} & \multirow{2}{*}{\makecell[c]{ASR\\ Network}} & \multirow{2}{*}{\makecell[c]{ASR\\ Train Data}}& \multicolumn{6}{c}{SNR} & \multirow{2}{*}{Avg} \\
         \cmidrule(lr){6-11}
         &  & & & & -6 dB & -3 dB & 0 dB & 3 dB & 6 dB & 9 dB & \\
         \hline
         \multirow{7}{*}{\rotatebox{90}{ADT}} & Unproc. & -  & WRConformer & Clean & 97.29 & 89.49 & 70.75 & 46.11 & 27.53 & 15.64 & 57.80\\
         & 1 \cite{wang_bridging_2019} & -   & WRBN & Noisy-160k & 54.03 & 31.52 & 17.90 & 10.83 & 7.16 & 4.93 & 21.06 \\
         & 2 & -  & WRConformer & Noisy-160k & 37.15 & 19.82 & 10.69 & 6.08 & 4.22 & 3.53 & 13.58\\
         & 3 & -  & WRConformer & Noisy-320k & 31.96 & 16.79 & 8.96 & 5.39 & 4.08 & 3.03 & 11.70 \\
         & 4 \cite{wang_bridging_2019} & GRN & WRBN & Enh.-160k & 39.36 & 20.86 & 11.02 & 6.16 & 4.40 & 3.20 & 14.16 \\
         & 5 & ARN (PCM) & WRConformer & Clean & \textbf{18.40} & 9.95 & \textbf{5.43}  & \textbf{3.83} & 3.39 & \textbf{2.88} & \textbf{7.31} \\
         & 6 & ARN (STOI) & WRConformer & Clean & 18.81 & \textbf{9.75} & \textbf{5.43} & 3.95 & \textbf{3.25} & 2.97 & 7.36\\
         \specialrule{1pt}{1pt}{0pt}
         
    \end{tabular}}
\end{table*}

\begin{table*}[!htbp]
    \centering
    \caption{ASR ($\%$WER) Results of the Proposed DR-ARN System and Comparison Systems on WSJ. All ASR Models Are WRConformer AM Trained on Clean Speech.}
    \label{tab:asr_wsj_drarn}
    \centering
    \scalebox{1}{
    \begin{tabular}[width=\linewidth]{ l l l  c  c c c c }
        \specialrule{1pt}{0pt}{4pt}
         \multirow{2}{*}{Row} & \multirow{2}{*}{\makecell[c]{SE\\Network}}  & \multirow{2}{*}{\makecell[c]{ASR\\ Train Data}} & \multicolumn{4}{c}{T60} & \multirow{2}{*}{Avg} \\
         \cmidrule(lr){4-7}
         & &  & 0.2--0.4 s & 0.4--0.6 s & 0.6--0.8 s & 0.8--1.0 s& \\
         \hline
          Unproc. & - & Anechoic & 15.21 & 33.55 & 50.36 & 57.93& 39.26\\
         1 & - &  Reverberant-160k & 3.47& 4.32& 4.52& 5.32 & 4.41 \\
         2 & - &  Reverberant-320k & 3.27 & 4.35 & 4.46 & 5.14 & 4.31 \\
         3 & ARN (PCM) &  Anechoic & 2.45 & 2.97 & 3.23 & \textbf{3.29} & \textbf{2.99} \\
         4 & ARN (STOI)  & Anechoic& \textbf{2.41} & \textbf{2.93} & \textbf{3.18} & 3.53 & 3.01 \\
         
         \specialrule{1pt}{0pt}{0pt}
         
    \end{tabular}}
\end{table*}

As the baselines for WSJ and CHiME-2 do not see the same amount of noise as the frontend training, we add two dynamic baselines on LibriSpeech to ensure that the E2E WRConformer trained on noisy speech observes the same amount of noise as that seen by the frontend during training. The DN-ARN takes 4 s long utterances during training, resulting in 312 hr noisy speech per epoch. In contrast, the E2E WRConformer sees 960 hr of noisy speech per epoch. Thus, for the first dynamic baseline, we generate noisy speech dynamically to train E2E WRConformer for 35 epochs, exactly matching the amount of noise seen by DN-ARN. For the second dynamic baseline, we train it in the same way as the first dynamic baseline but add clean speech to the training set to create a multi-conditioned training set. The setup for dynamic baselines is the same as for E2E WRConformer on clean speech.

\begin{table*}[htbp!]
    \centering
    \caption{ASR ($\%$WER) Results of the Proposed NR-ARN System and Comparison Systems on WSJ. Clean Data Denotes Anechoic Clean and Noisy Data Denotes Reverberant Noisy. All ASR Models are WRConformer AM. ASR Training Data is Anechoic in the Clean Case and Reverberant in the Noisy Case.}
    \label{tab:asr_wsj_nrarn}
    \centering
    \scalebox{1}{
    \begin{tabular}[width=\linewidth]{ c c  l  l  c  c c c c c c }
        \specialrule{1pt}{0pt}{4pt}
         \multirow{2}{*}{\makecell[c]{Test\\ Noise}} & \multirow{2}{*}{\makecell[c]{Test\\ T60}}&\multirow{2}{*}{\makecell[c]{SE\\ Network}}  & \multirow{2}{*}{\makecell[c]{ASR\\ Train Data}} & \multicolumn{6}{c}{SNR} & \multirow{2}{*}{Avg} \\
         \cline{5-10}
         && & & -6 dB & -3 dB & 0 dB & 3 dB & 6 dB & 9 dB & \\
         \hline
         \multirow{20}{*}{\rotatebox{90}{ADT}} &  \multirow{6}{*}{0.2--0.4 s}& - & Clean & 96.02 & 93.01 & 84.49 & 67.93 & 49.65 & 35.31 & 71.07\\
         & & - & Noisy-160k & 52.39	&31.38&	17.65	&10.24&	6.52&	5.02	&20.53 \\
         & & -  & Noisy-320k  & 48.60&	28.18	&15.57	&9.27	&6.34&	4.89	&18.81\\
         & & ARN (PCM)  & Clean & 37.97 & \textbf{20.81} & 12.03 & 7.72 & 5.35 & \textbf{4.12} & 14.66\\
         & & ARN (STOI) & Clean & \textbf{37.96} & 20.90 & \textbf{11.42} & \textbf{7.17} & \textbf{5.13} & \textbf{4.12} & \textbf{14.45}\\
         \cline{2-11}
         
         &  \multirow{6}{*}{0.4--0.6 s}& -&  Clean & 96.70 & 94.98 & 89.45 & 79.14 & 66.70& 53.64 & 80.10\\
         & & - & Noisy-160k & 55.33&	34.73&	19.81	&12.08&	8.04&	5.97	&22.66 \\
         & & -  & Noisy-320k  &  51.58&	31.38&	18.56&	11.28&	7.41 &5.58&	20.96\\
         & & ARN (PCM)  & Clean & 46.54 &	27.73 &	15.09 &	9.38 & 5.97& 4.69&18.23 \\
         & & ARN (STOI)  & Clean& \textbf{45.98} & \textbf{27.37} & \textbf{15.01} & \textbf{9.36} & \textbf{5.95} & \textbf{4.65} & \textbf{18.05}\\
         
          \cline{2-11}
         &  \multirow{6}{*}{0.6--0.8 s}& - & Clean & 96.90&	95.75	&91.89&	85.16&	75.62	&66.62	&85.32\\
         & & - & Noisy-160k & 58.86	&37.87&	21.93	&13.97	&9.48	&6.96 & 24.84 \\
         & & -  & Noisy-320k & 55.46	&34.40	&20.54&	12.81&	8.26&	6.02&	22.91 \\
         & & ARN (PCM)  & Clean & 52.74 &	32.27&	18.37&	11.31&	7.44&	5.62	&21.29 \\
         & & ARN (STOI)  & Clean&  \textbf{52.58}&	\textbf{31.06}&	\textbf{17.86}	&\textbf{10.91}	&\textbf{7.21}&	\textbf{5.48} &	\textbf{20.85}\\
         
          \cline{2-11}
         &  \multirow{6}{*}{0.8--1.0 s}& Mixture & Clean & 97.13&	96.25&	93.88	&88.58&	81.80	&74.22	&88.64 \\
         & & - & Noisy-160k & 61.04	&39.64	&24.64&	15.36	&10.21	&7.43&	26.38 \\
         & & -  & Noisy-320k  & 58.48	&37.06	&21.94&	13.70&	9.30&	7.19	&24.61\\
         & & ARN (PCM)  & Clean & 57.55&	34.77&	20.36	&12.23	&\textbf{8.31}	&\textbf{6.16}&	23.23 \\
         & & ARN (STOI)  & Clean&  \textbf{57.06} &	\textbf{34.46} &	\textbf{20.04} &	\textbf{12.11} &	8.32	&6.19&	\textbf{23.03} \\
        \specialrule{1pt}{0pt}{0pt}

    \end{tabular}}
\end{table*}

\subsection{Cross-corpus Generalization to CHiME-4}
To validate the cross-corpus generalization of the proposed system, we take the CHiME-4 \cite{vincent2017analysis} corpus for evaluation, which comprises simulated and real (recorded) noisy speech. There are 1640 real and 1640 simulated noisy utterances from 4 different speakers in the development set. The test set consists of 1320 real and 1320 simulated noisy utterances from 4 other speakers. Real data is recorded speech in noisy environments (bus, cafe, pedestrian area, and street junction) uttered by real talkers, and simulated data is generated by mixing recorded speech via a close-talking microphone and separately recorded noises. The training set is not used in this study. We select the fifth channel for single-channel track evaluation. In our decoupled system, we take the ARN trained on LibriSpeech with STOI validation as the frontend, and WRConformer AM trained on WSJ0 SI-84 as the backend. We follow the same procedure as in \cite{yang2022conformer} for LM rescoring and unsupervised speaker adaptation.

\section{Results and Discussions}\label{sec:result}
This section presents and analyzes the evaluation and comparison results of the decoupled systems and the corresponding baselines on three corpora. As the focus of this study is on ASR performance, we only provide a summary of the SE performance for each system.

\subsection{Results on WSJ}\label{sec:result_dnarn_wsj}
\subsubsection{Denoising Only}
On average, DN-ARN with PCM and STOI validation improves STOI by $19.22\%$ and $19.25\%$, and PESQ by $1.38$ and $1.39$, respectively, across different SNRs. DN-ARN with STOI validation slightly outperforms DN-ARN with PCM validation.

ASR evaluation results are shown in Table~\ref{tab:asr_wsj_dnarn}. Results on babble and cafeteria noise are averaged and denoted as ADT. When tested on the WSJ anechoic-clean utterances, the WERs for the WRConformer AM trained on clean, noisy-160k, and noisy-320k are $2.04\%$, $2.37\%$, and $2.05\%$, respectively. We first evaluate and compare AMs trained on noisy speech. The results in row 2 and row 3 outperform those of row 1 by $35.5\%$ and $44.4\%$, respectively. The results demonstrate the effectiveness of the noise-independent WRConformer AM baselines.

We next evaluate the decoupled system and compare it with the WRBN-based system that is trained on GRN (gated residual network) \cite{tan2018gated} enhanced speech \cite{wang_bridging_2019}. The decoupled system in row 5 achieves $7.31\%$ WER on average, outperforming that in row 4 by $48.4\%$. The decoupled system with DN-ARN with STOI validation performs closely in row 6. Both decoupled systems substantially outperform all baselines. These results demonstrate that SE translates to improved ASR results for anechoic-noisy speech.

\subsubsection{Dereverberation Only}

DR-ARN with PCM validation and STOI validation are evaluated on reverberant-clean WSJ speech in four T60 ranges. On average the DR-ARN with PCM and STOI validation improves STOI by $17.49\%$ and $17.51\%$, respectively. The two DR-ARNs have the same PESQ scores in the four T60 ranges, and on average improve PESQ by $1.42$.

Table~\ref{tab:asr_wsj_drarn} provides the ASR results of the proposed system and baselines. Tested on the anechoic-clean WSJ utterances, the WRConformer AMs trained on anechoic, reverberant-160k, and reverberant-320k have $2.04\%$, $3.03\%$, and $3.05\%$ WER, respectively. The results in row 2 outperform those in row 1 by $2.3\%$ relatively, showing a slight effect of training data size. In rows 3 and 4, DR-ARN with PCM and STOI validation produces $2.99\%$ and $3.01\%$ WER, respectively. The results in row 3 outperform those of the row 2 by $30.6\%$. This shows that DR-ARN is capable of effective speech dereverberation, and improved dereverberation translates to better ASR with the corresponding AM trained on anechoic speech.

\begin{table*}[!htbp]
    \centering
    \caption{ASR ($\%$WER) Results of the Proposed DN-ARN System and Comparison Systems on CHiME-2. All ASR Models Are Trained on Reverberant Speech. `Enh.' and `Feat.' Denote `Enhanced' and `Feature', Respectively.}
    \label{tab:asr_chime2_dnarn}
    \centering
    \scalebox{1}{
    \begin{tabular}[width=\linewidth]{ l l  l  l  c  c c c c c c}
        \specialrule{1pt}{0pt}{4pt}
         \multirow{2}{*}{Row} & \multirow{2}{*}{\makecell[c]{SE\\ Network}} & \multirow{2}{*}{\makecell[c]{ASR\\ Network}} & \multirow{2}{*}{\makecell[c]{ASR\\ Train Data}} & \multicolumn{6}{c}{SNR} & \multirow{2}{*}{Avg} \\
         \cmidrule(lr){5-10}
         & & & & -6 dB & -3 dB & 0 dB & 3 dB & 6 dB & 9 dB & \\
         \hline
         Unproc.& - & WRConformer & Clean & 73.81 & 64.88  & 57.59 & 45.10 & 36.05 & 28.54 & 51.00 \\
         1 \cite{wang_bridging_2019} & -  & WRBN & Noisy-160k & 17.45 & 13.06 & 10.69 & 8.82 & 7.72  & 6.63 & 10.73 \\
         2 & -& WRConformer & Noisy-160k & 19.82 & 13.30 & 10.91 & 9.25 & 7.25 & 6.63 & 11.19\\
         3 & - & WRConformer &  Noisy-320k & 16.25 & 10.16 & 9.25 & 7.08 & 6.54 & 5.64 & 9.15\\
         4 \cite{wang_bridging_2019}& - & WRBN & CHiME-2 Noisy & 14.83 & 9.98 & 8.95 & 6.78 & 6.26 & 5.49 & 8.72 \\
         5 & - & WRConformer & CHiME-2 Noisy & 14.44 & 10.33 & 7.92 & 6.73 & 6.03 & 5.49 & 8.49\\
         \hline
         6 \cite{wang_bridging_2019} & GRN & WRBN & Enh.-160k  & 15.45 & 11.04 & 9.70  & 7.10 & 6.54 & 5.51 & 9.22\\
         7 \cite{wang_enhanced_2019}  & GRN & WRBN & Enh. Feat.-160k & 13.11 & 9.43 & 7.92 & 6.20 & 5.45 & 4.54 & 7.78\\
         8 \cite{plantinga2021perceptual} & Wide ResNet  & MLP & Multi-conditioned & 15.2 & 10.9 & 8.3 & 6.7 & 5.8 & 5.2 &8.7  \\
         9 & ARN (PCM) &  WRConformer & Clean & 13.30 & 9.66 & 7.83 & 6.41 & 5.25 & 4.60 & 7.84 \\
         10 & ARN (STOI) & WRConformer & Clean & 9.94 & 7.04 & 6.50 & 5.45 & 4.54 & 4.22 & 6.28\\
         11 & CrossNet (PCM) & WRConformer & Clean &8.46 & \textbf{6.54} & 5.53 & \textbf{4.84} & 4.48 & \textbf{4.09} & 5.62 \\
         12 & CrossNet (STOI) & WRConformer & Clean & \textbf{7.75} & 6.74 & \textbf{5.16} & 5.21 & \textbf{4.37} & 4.18 & \textbf{5.57} \\
         13 & CrossNet (SI-SDR) & WRConformer & Clean & 9.14 & 7.53 & 6.22 & 5.44 & 5.23 & 4.61 & 6.36 \\

         \specialrule{1pt}{0pt}{0pt}
    \end{tabular}}
\end{table*}

\subsubsection{Denoising and Dereverberation}
NR-ARN is evaluated on babble and cafeteria noise at six SNR levels in four T60 ranges. NR-ARN with PCM validation performs similarly to that with STOI validation. In the T60 range of [0.8, 1.0] s, NR-ARN with STOI validation yields $28.47\%$ and $27.64\%$ improvement in STOI and $1.05$ improvement in PESQ, for babble and cafeteria noise, respectively. The results illustrate the ability of NR-ARN in joint denoising and dereverberation.



Table~\ref{tab:asr_wsj_nrarn} presents the ASR results of the proposed decoupled system of NR-ARN and comparisons with baselines. Results on babble and cafeteria noise are averaged and denoted as ADT. Boldface entries indicate the best results for each T60 range. When tested on the WSJ anechoic-clean utterances, the WER for the WRConformer trained on clean, noisy-160k, and noisy-320k are $2.04\%$, $3.57\%$, and $3.12\%$, respectively. AM with NR-ARN with STOI validation outperforms that with PCM validation and the noise- and reverberation-independent baselines in all test conditions. For four T60 ranges in increasing order, AM with NR-ARN with STOI validation outperforms the baseline trained on noisy-320k by $23.2\%$, $13.9\%$, $9.0\%$, and $6.4\%$, respectively. These results demonstrate that NR-ARN enhanced speech results in improved ASR performance for reverberant-noisy speech.

\subsection{Results on CHiME-2}\label{sec:result_dnarn_chime2}
On average, DN-ARN with PCM and STOI validation improves STOI by $12.59\%$ and $12.07\%$, and PESQ by $1.04$ and $1.02$, respectively. For CrossNet with PCM, STOI, and SI-SDR validation, the STOI improvements are $9.40\%$, $9.02\%$, and $9.09\%$ and PESQ improvements are $0.75$, $0.73$, and $0.75$, respectively.

ASR results are presented in Table~\ref{tab:asr_chime2_dnarn}. When tested on CHiME-2 reverberant-clean utterances, the WERs for the WRConformer AM trained on clean, CHiME-2 noisy, noisy-160k, and noisy-320k are $2.90\%$, $4.35\%$, $3.48\%$, and $3.26\%$, respectively.


\begin{table*}[!htbp]
    \centering
    \caption{ASR ($\%$WER) Results of the Proposed DN-ARN and Comparison Systems on LibriSpeech. All ASR Models Use E2E WRConformer.}
    \label{tab:asr_librispeech_dnarn}
    \centering
    \scalebox{1}{
    \begin{tabular}[width=\linewidth]{ c c l l l  c  c c c c c c }
        \specialrule{1pt}{0pt}{4pt}
         \multirow{2}{*}{\makecell[c]{Test\\ Noise}} & \multirow{2}{*}{\makecell[c]{Test\\ Data}}& \multirow{2}{*}{Row} &\multirow{2}{*}{\makecell[c]{SE\\Network}} &  \multirow{2}{*}{\makecell[c]{ASR\\ Train Data}}& \multicolumn{6}{c}{SNR} & \multirow{2}{*}{Avg} \\
         \cmidrule(lr){6-11}
         & &  & & & -5 dB & -2 dB & 0 dB & 2 dB & 5 dB & 10 dB &\\
         \hline
         \multirow{12}{*}{\rotatebox{90}{\makecell[c]{ADT}}} & \multirow{6}{*}{\rotatebox{90}{\makecell[c]{test\\-clean}}} & Unproc. & -   & Clean & 100.4 & 88.7 &  71.5 & 48.5 & 20.2 & 5.1 & 55.7 \\
          & & 1 & -   & Noisy-4k-hr & 41.1 & 20.3 & 13.1 & 9.4 & 6.8 & 6.3 & 16.1 \\
          & & 2 & -   & Dynamic Noisy & 39.0 & 17.7 & 10.8 & 7.1 & 4.6 & 3.2 & 13.7 \\
          & & 3 & -   & Dynamic Multi-conditioned& 55.3 & 29.7 & 19.2 & 13.4 & 9.1 & 6.5 & 22.2\\
          & & 4 & ARN (PCM)  & Clean & 28.7 & \textbf{13.0} & \textbf{7.9} & 5.3 & \textbf{3.5} & \textbf{2.4} & \textbf{10.1}\\
          & & 5 & ARN (STOI)  & Clean & \textbf{28.6} & \textbf{13.0} & \textbf{7.9} & \textbf{5.2} & \textbf{3.5} & \textbf{2.4} & \textbf{10.1}\\
        \cline{2-12}    
                  & \multirow{6}{*}{\rotatebox{90}{\makecell[c]{test\\-other}}} & Unproc. & -   & Clean & 102.8 & 97.6 & 87.3 & 71.3 & 44.4 & 15.7  & 69.8\\
          & & 6 & -   & Noisy-4k-hr & 63.9 & 41.1 & 29.8 & 22.0 & 15.3 & 12.0 & 30.7 \\
          & & 7 & -   & Dynamic Noisy & 62.2 & 38.4 & 27.2 & 18.8 & 12.4 & 8.00 & 27.9 \\
          & & 8 & -   & Dynamic Multi-conditioned& 77.3 & 52.7 & 39.6 & 29.9 & 20.9 & 14.1 & 39.3 \\
          & & 9 & ARN (PCM)  & Clean & 52.4 & 31.0 & \textbf{21.3} & 14.9 & 9.7 & \textbf{6.2} & 22.6\\
          & & 10 & ARN (STOI)  & Clean & \textbf{52.1} & \textbf{30.7} & 21.4 & \textbf{15.0} & \textbf{9.8} & \textbf{6.2} & \textbf{22.5}\\
        \specialrule{1pt}{0pt}{0pt} 
            \end{tabular}}
\end{table*}

Trained on noisy speech directly, the baseline in row 3 improves WER by $14.7\%$ compared with those of row 1. The noise-dependent WRConformer AM in row 5 outperforms the noise-dependent WRBN in row 4 by $2.6\%$. WRBN-based models perform consistently worse than WRConformer AMs, demonstrating the effectiveness of the WRConformer baselines.

We next evaluate the proposed decoupled system and compare it with the other systems that incorporate SE. Row 6 of Table~\ref{tab:asr_chime2_dnarn} is WRBN trained on GRN enhanced speech \cite{wang_bridging_2019}, and row 7 is WRBN trained on GRN enhanced magnitude spectra \cite{wang_enhanced_2019}. A perceptual loss based model \cite{plantinga2021perceptual} is included in row 8, and it employs a Wide ResNet trained with a perceptual loss as the frontend and its backend use an off-the-shelf Kaldi CHiME-2 recipe trained on multi-conditioned speech (WSJ0 SI-84 clean + CHiME-2 reverberant-clean + CHiME-2 reverberant-noisy). The proposed system with DN-ARN with STOI validation achieves $6.28\%$ WER, which outperforms the previous best \cite{wang_enhanced_2019} by $19.3\%$ relatively. In row 12, the decoupled system with CrossNet with STOI validation achieves a $5.57\%$ WER, outperforming the previous best by $28.3\%$ relatively. Also, STOI validation produces better results than PCM or SI-SDR validation. These results clearly demonstrate that either the time or T-F domain frontend can deal with reverberant-noisy speech for decoupled robust ASR, expanding our previous findings on time-domain frontend only \cite{yang2022time}.

\subsection{Results on LibriSpeech}

As in the evaluation on WSJ, the decoupled system with DN-ARN on LibriSpeech is evaluated with babble and cafeteria noises. Results are averaged across two noise types and denoted as ADT in Table~\ref{tab:asr_librispeech_dnarn}. On the test-clean set, DN-ARN with PCM and STOI validation improves STOI by $21.47\%$ and $21.51\%$, respectively, and PESQ by $1.45$ with both validations. On the test-other set, DN-ARN with PCM and STOI validation shows improvements in STOI by $20.50\%$ and $20.62\%$, and in PESQ by $1.20$ and $1.22$, respectively.

For ASR performance, the WERs of each system tested on clean speech are presented in Table~\ref{tab:asr_libri_clean}. The E2E WRConformer trained on clean speech outperforms all baseline models. The E2E WRConformer trained on dynamic noisy speech outperforms the other two baselines. The WERs of the ASR model trained on dynamic multi-conditioned speech are the highest among all the baselines, due to its training to match the data observed for SE frontend training which is quite different from clean test data.

\begin{table}[!htbp]
    \centering
    \caption{ASR ($\%$WER) Results of Different E2E WRConformer Models Tested on LibriSpeech Test Sets. `Dyn.' and `cond.' Denote `Dynamic' and `conditioned', respectively.}
    \label{tab:asr_libri_clean}
    \centering
    \resizebox{0.99\linewidth}{!}{
    \begin{tabular}[width=\linewidth]{ c  c c c c }
        \specialrule{1pt}{0pt}{4pt}
        \multirow{2}{*}{Train Data} & \multicolumn{4}{c}{Test Data}\\
        \cmidrule{2-5}
         & dev-clean & dev-other & test-clean & test-other \\
         \hline
          Clean & 1.7 & 3.9 & 1.9& 4.1\\
         Noisy-4k-hr&4.3 & 8.4 & 4.6 & 8.5 \\
          Dyn. Noisy&2.3 & 5.5& 2.6 & 5.7 \\
          Dyn. Multi-cond. & 6.0 & 10.7 & 6.1 & 10.7 \\
         \specialrule{1pt}{0pt}{0pt}
    \end{tabular}}
\end{table}

Tested under ADT noises, the proposed decoupled systems outperform all baselines. The decoupled systems with DN-ARN with PCM and STOI validation have very close performances. The system with STOI validation in row 5 and row 10 of Table~\ref{tab:asr_librispeech_dnarn} outperforms the best baseline trained on dynamic noisy speech in row 2 and row 7 by $26.3\%$ and $19.4\%$ relatively on test-clean and test-other datasets, respectively. This demonstrates that, using the same amount of noisy speech, training an SE frontend is more effective than training an ASR model for robust ASR.

\subsection{Results on CHiME-4}
Table~\ref{tab:asr_chime4} presents CHiME-4 evaluation results, as well as comparisons with many baseline systems. The comparison systems include six trained on CHiME-4 noisy speech directly \cite{chen2018building, guo2021recent, jahn2016wide, du2016ustc, wang2020complex, yang2022conformer}, and five that leverage additional speech materials for pretraining and employ CHiME-4 training data for fine-tuning \cite{zhu2022joint, yang2023fat, hu2023wav2code, wang2022improving, chang2022end}. Withouts training on CHiME-4, the proposed decoupled system achieves $3.32/4.44\%$ WER on simulated/real test data. Our system on the simulated data dramatically outperforms the previous best by $45.8\%$, utilizing around 100 times less cross-corpus speech data. With the same amount of cross-corpus speech material, the decoupled system outperforms the previous best \cite{hu2023wav2code} on the real data by $16.2\%$. It is worth noting that all of the previous state-of-the-art methods on the single-channel CHiME-4 evaluation utilize CHiME-4 training data, making ours the first state-of-the-art method that benefits from pretrained speech enhancement on a different corpus and ASR. Another interesting observation regards the relative WER scores between the simulated and real data: the trend of our system reverses that of all the other baselines. It seems more reasonable to expect better results on simulated data, as exhibited in our system, and this trend is more consistent with two-channel and six-channel CHiME-4 evaluation results \cite{wang2020complex}. This demonstrates that the decoupled system generalizes well to cross-corpus real acoustic scenarios.

\begin{table}[htbp!]
    \centering
    \caption{ASR (\%WER) Results of The Proposed and Comparison Systems on CHiME-4 (Single-channel). `Iter. Spk. Adapt.' denotes `Iterative Speaker Adaptation'.}
    \label{tab:asr_chime4}
    \centering
    \resizebox{0.99\linewidth}{!}{
    \begin{tabular}[width=\linewidth]{ l p{0.25\linewidth}<{\centering}  c  c  c  c }
         \specialrule{1pt}{0pt}{0pt}
         
          \makecell[l]{\multirow{3}{*}{System}} & \multirow{3}{*}{\makecell[c]{Cross-corpus\\Speech Hour (hr) / \\  Train on CHiME-4?}} &\multicolumn{2}{c}{\multirow{2}{*}{Dev. Set}} &  \multicolumn{2}{c}{\multirow{2}{*}{Test Set}} \\
          \\
          \cmidrule(lr){3-4}
          \cmidrule(lr){5-6}
         & & Simu. & Real & Simu. & Real \\
            \hline
          Kaldi Baseline \cite{chen2018building} & \multirow{6}{*}{0 / \ding{51}} & 6.81 & 5.58 & 12.15  & 11.42 \\
           ESPnet Conformer \cite{guo2021recent} & & 9.10 & 7.90 & 14.20 & 13.40 \\
          WRBN \cite{jahn2016wide} & & 6.69 & 5.19 & 11.11 & 9.34\\
         Du \emph{et al.} \cite{du2016ustc}& & 6.61 & 4.55 & 11.81 & 9.15 \\
         Wang \emph{et al.} \cite{wang2020complex}& & 4.99 & 3.54 & 9.41 & 6.82 \\
         Yang \emph{et al.} \cite{yang2022conformer} & & 4.99 & 3.35 & 8.61 & 6.25 \\
         
        \hline
          Zhu \emph{et al.} \cite{zhu2023robust} & 1k / \ding{51} & - & 3.1 & - & 5.8 \\
          Yang \emph{et al.} \cite{yang2023fat} & 1k / \ding{51} & - & 3.1 & - & 5.7\\
          Hu \emph{et al.} \cite{hu2023wav2code} & 1k / \ding{51} & - & 2.6 & - & 5.3 \\
          Wang \emph{et al.} \cite{wang2022improving} & 60k / \ding{51} & - & 2.7 & - & 5.5 \\
        Chang \emph{et al.} \cite{chang2022end} & 94k / \ding{51} & 3.16 & \textbf{2.03} & 6.12 & \textbf{3.92} \\
           \hline
         Enh. + 3-gram & \multirow{4}{*}{1k / \ding{55}}  & 6.95 & 7.63 &7.32  & 10.59 \\
         
         \hspace{1mm} + 5-gram + RNNLM & & 5.13 & 5.43 & 5.63 & 7.83 \\
         
         \hspace{1mm} + 5-gram + LSTMLM & & 3.85 & 4.21 & 4.23 & 6.74\\
        

         \hspace{2mm} + Iter. Spk. Adapt. & & \textbf{3.09} & 2.98 & \textbf{3.32} & 4.44   \\

          \specialrule{1pt}{0pt}{0pt}
         
    \end{tabular}}
\end{table}

\section{Concluduing Remarks}\label{sec:conclusion}
This study aims to eliminate the divide between speech enhancement frontend and recognition backend in monaural robust ASR. The time-domain ARN and T-F domain CrossNet are employed as SE frontends to WRConformer based ASR models trained on clean speech only. The proposed system fully decouples SE and ASR. Results on the WSJ, CHiME-2, LibriSpeech, and CHiME-4 corpora on denoising, dereverberation, or both, show that SE gains translate to ASR gains. The proposed system outperforms the interference-independent baselines for all test conditions. Our CHiME-2 results have updated the previous best WER by $28.3\%$ relatively in the standard evaluation, and we achieve $3.32/4.44\%$ WER on simulated/real data without training on CHiME-4. These CHiME-4 results cut the previous best WER on simulated data by a large margin, and represent the best WER on real data using the same amount of out-of-corpus speech materials.

The largest STOI and PESQ improvements as well as the ASR improvements come from DN-ARN on WSJ, showing that ARN excels in mapping anechoic-noisy speech into anechoic-clean speech, the task that ARN is originally designed for \cite{pandey2022self}. For reverberant and reverberant-noisy speech, the proposed system using DR-ARN, NR-ARN, and CrossNet performs very well.

The model selection criterion of maximum STOI tends to produce better ASR performance than PCM validation, especially for reverberant-noisy speech. For DN-ARN and DR-ARN on WSJ, the system with PCM validation outperforms the system with STOI validation. This gets reversed for DN-ARN on CHiME-2, CrossNet on CHiME-2, and NR-ARN on WSJ, where input speech to ASR is reverberant-noisy. But the WER improvements in DN-ARN on CHiME-2 and NR-ARN on WSJ are larger than the slight improvements in DN-ARN and DR-ARN on WSJ, and CrossNet on CHiME-2. Consistent with \cite{moore2017speech}, our observations suggest that STOI validation is a strong criterion for SE model selection for downstream ASR tasks. If an ASR model trained on clean speech is available, another possibility is to perform SE model selection directly using ASR accuracy, which is expected to better match the ASR task. Further research is needed to compare ASR and STOI criteria.

In future work, we plan to improve the performance of the frontend for joint denoising and dereverberation. In addition, we plan to extend our approach to multi-talker speaker separation to test its ability to eliminate interfering speakers, and multi-channel robust ASR tasks.

\section{Acknowledgement}
The authors would like to thank Vahid Ahmadi Kalkhorani for sharing the code of CrossNet.

\bibliographystyle{IEEEtran}
\bibliography{IEEEabrv, yyf_refs}

\end{document}